\documentclass[a4paper]{article}
\usepackage{epsfig,amsmath,amsfonts,amssymb,setspace,multirow,textcomp}
\usepackage[T1]{fontenc}
\usepackage[english]{babel}
\makeatletter
\renewcommand{\@oddhead}{\textit{Advances in Astronomy and Space Physics} \hfil}
\renewcommand{\@evenfoot}{\hfil \thepage \hfil}
\renewcommand{\@oddfoot}{\hfil \thepage \hfil}
\makeatother

\renewenvironment{thebibliography}[1]{\begin{oldthebibliography}{#1}\setlength{\parskip}{0ex}\setlength{\itemsep}{0ex}}{\end{oldthebibliography}}
\begin{document}

\fontsize{11}{11}\selectfont % the font size cannot be changed in any case!
%  insert your title, authors information and text instead of the one provided below
\title{Prospects for gamma-ray observations of Hercules cluster}
\author{\textsl{V.\,V.~Voitsekhovskyi$^{1}$}}
\date{\vspace*{-6ex}}
\maketitle
\begin{center} {\small $^{1}$Taras Shevchenko National University of Kyiv,
Glushkova ave., 4, 03127, Kyiv, Ukraine\\
{\tt v.v.voitsekhovskyi@gmail.com}}
\end{center}

\begin{abstract}

\indent \indent Galaxy clusters (GCs) are the largest and most massive
gravitationally bound objects in the large-scale
structure of the Universe. Due to keV temperatures of virialized gas
in the intracluster medium (ICM)
and presence of cosmic rays (CRs), GCs are effective sources of
thermal X-ray radiation and non-thermal
leptonic (synchrotron) radio emission. GCs are also store-rooms
for hadronic CRs, but  non-thermal
hadronic
gamma-ray emission (mainly, due to pp collisions and subsequent pion decay)
from GCs has not been detected yet.
In this work we simulate the expected non-thermal hadronic gamma-ray and neutrino  emission
from dominant part of Hercules cluster - GC A2151 -  and estimate a perspective of detection
of this emission by existing   (Fermi-LAT, LHASSO, IceCube)
and planned (CTA, IceCube-Gen2) ground-based ans space-based detectors.

\end{abstract}

\section*{\sc introduction}

Galaxy clusters (GCs)  are the largest virialized structures in the
Universe at the present cosmological time $t_0=13.8$ Gyr (a redshift $z_0=0.0$)\cite{Voit_04}.
A growth of initial density fluctuations due to gravitational instability led to the
formation of main elements of large scale structure: sheet-like structures, bordered
low density voids, and  filaments,  connected to the high-density nodes - galaxy clusters
and superclusters \cite{Springel_Frenk_White_06}. The total gravitational masses
$M_{cl} \sim 10^{14}-10^{15}$ M$_{\odot}$ of  GCs include
dominant contribution of dark matter (about 80 \% of mass) and much smaller contribution
of baryonic matter ( $\sim 15\% $ of mass in intracluster medium (ICM) gas and
$ \sim 5 \% $ of mass in stars/galaxies) \cite{Mohr_etal_99},
\cite{Berezinsky_Blasi_Ptuskin_97}, \cite{Allen_Evrard_Mantz_2011}.
Virial temperatures of ICM plasma
$k_BT\sim GM_{cl}m_p/R_{cl}$ in GCs with typical size $R_{cl}\sim 2-3$ Mpc are of
the order of $1-10$ keV, therefore GCs
are luminous sources of thermal  X-ray emission, with typical luminosities
$L_X \sim 10^{44}-10^{46}$ erg s$^{-1}$.

Formation of gravitationally bound and virialised GCs are long-lasting violent process of main halo
collapse and substructures' merger, that is often not completed yet. Most clusters have not yet fully
relaxed and accretion and merger processes are still ongoing.  Magnetohydrodynamic (MHD) flows in
ICM  and in active galactic nuclei (AGN), rich in shock waves, efficiently accelerate
cosmic rays (CR) and enhance/generate magnetic fields in ICM. Once more, in observed
$B_{cl}\sim 1-10\,\mu$G GC magnetic fields the time of diffusive escape of CR from GCs
$t_{dif}\sim R^2_{cl}/D(E)$ is of order or larger than the age of Universe for typical value of
diffusion coefficient $D(E)\sim 10^{28}(E/10 GeV)^{0.5}$cm$^2$ s$^{-1}$. As a result,  GCs
are effective store-rooms for accelerated  CR \cite{Berezinsky_Blasi_Ptuskin_97} and, therefore,
GCs are expected to be promising sources of non-thermal leptonic (synchrotron) radio emission
as well as of hadronic (mainly, due to pp collisions and  subsequent neutral pion decay)
\cite{Pfrommer_Eblin_04}, \cite{Pinzke_Pfrommer_10}.

%Indeed,  a non-thermal synchrotron radio emission
%is widely presented in observations of GCs

Indeed, a non-thermal synchrotron radio emission

in MHz $-$ GHz  range
from relativistic electrons ($E_e\sim 10$ GeV)  in  clusters'
magnetic fields
($\nu_{syn}\sim 4(B_{cl}/1\mu$ G)($E_e$ /10 GeV)$^2$ GHz)
is widely presented in observations of GCs \cite{van_etal_19},
but  non-thermal  gamma-ray emission from GCs has not been detected yet \cite{Ackerman_etal_14},
\cite{Wittor_21}.

Both  non-thermal radio- and thermal  X-ray emission, supplemented by data of {\it Planck}
thermal SZ effect, give a valuable information about dark/baryon matter distribution and physical
processes in ICM of GCs \cite{Adam_etal_20}. In our work
we have carried out a modelling of non-thermal hadronic gamma-ray and neutrino  emission
from Abell cluster A2151 (a dominant part of Hercules cluster) and estimate prospects for gamma-ray
observations of A2151 by existing and planned ((Fermi-LAT, LHASSO, IceCube, CTA, IceCube-Gen2)
ground-based and space-based detectors. For calculation we use code  MINOT \cite{Adam_etal_20}  and  $\Lambda$CDM cosmological model with $ H_0=70 $ km s$^{-1}$ Mpc$^{-1}$, $\Omega_M=0.3$,  $\Omega_{\Lambda}=0.7$.

\section*{\sc hercules cluster}

The nearby ( z=0.0367) Hercules supercluster is one of the most massive structures in the Local Universe.
It is composed by Abell clusters  A2151 (the Hercules cluster),  A2147 and A2152 \cite{Barmby_Huchra_98}.
The most prominent GC A2151 with mass $ M_{200}= 4.00\times 10^{14}M_{\odot}$ and  size $R_{200} = 1.45$ Mpc
includes  three  subclusters and is still in the processes of collapsing and mergers \cite{Agulli_etal_17}.
There are 7 AGN-type galaxies in central part of A2151, including NGC 6050 in  Arp 272 - a pair of interacting
galaxies NGC 6050  and IC 1179.

Considered in this paper GC A2151 contains nearly 360 galaxies, and its optical center have coordinates
R.A. = $241.34^{\circ}$ and Dec = $17.75^{\circ}$ \cite{Agulli_etal_17}. This cluster demonstrates
significant division into several substructures, particularly, bright central bimodal subclump A2151C
and two fainters - A2151E and A2151N. Based on Chandra and XMM-Newton observation, A2151C also can be
divided into two substructures - bright central A2151C(B) and fainter lateral A2151C(F) \cite{Tiwari_PalSingh_20}.
According to state of the art sensitivity of our $\gamma$-ray detectors, only A2151C has a chance to be detected,
so aim of this paper is to inspect this problem.

\indent \indent
Physical conditions and spherically-symmetric dark/baryonic matter distribution in ICM of GC
are determined mainly by two global GC parameters -  its total mass $M_{cl}$ and size $R_{cl}$.
 There are several types of characteristic masses that we have used in modelling. $M_{500}$ denotes
 the mass that is enclosed inside the  radius $R_{500}$, where a mean density (averaged on volume)
 exceeds critical density 500 times $\rho /\rho_{crit} = 500$. Corresponding values are related
 to each other by following relation:

 \begin{equation}
    M_{500} = \frac{4\pi}{3} 500\rho_{crit} R_{500}^3
    \label{eqn:MassRadius}
\end{equation}

Total mass $M_{tot}$ of GC can be expressed as $M_{tot} = M_{HSE}/(1-b_{HSE})$, where $M_{HSE}$ is
the hydrostatic mass derived from hydrostatical equilibrium and $b_{HSE} \sim 0.2$ is the
hydrostatic mass bias \cite{Planck_Col_14_HSE}, \cite{Piffaretti_Valdarnini_08}.
Spatial  distribution of gas mass fraction in  GC $f_{gas}(r)$ is determined as the ratio  of the gas mass $M_{gas}(r)$ and the total mass $M_{tot}(r)$ $f_{gas}(r)=M_{gas}(r)/M_{tot}(r)$.

Gas electron density profile $n_{e}(r)$ is an important parameter for estimating gamma-ray emission, because it is proportional to total gas density  $n_{gas}(r)$, that  describes the intensity of proton-proton interactions. There are several theoretical models to describe  $n_{e}(r)$, particularly, analytical beta model  \cite{Cavaliere_FuscoFemiani_78} that was used in this work:
\begin{equation}
    n_{e}(r) = n_{e,0} \left[ \, 1 + \left(\frac{r}{r_c} \right)^2 \, \right ]^{-3\beta_{dens} / 2}
    \label{eqn:beta}
\end{equation}
\noindent  where $n_{e,0}$ is the normalisation factor,  $r_c$ is the core radius of density distribution,  $\beta_{dens}$  is the density index.

Thermal gas pressure is also important, because it allows to normalize the number of CR.
Electron thermal pressure $P_e(r)$ is often  described by Generalized Navarro Frank White
(GNFW) profile \cite{gNFW_07} as an  improved version of the earlier Navarro Frenk White
profile (NFW) \cite{NFW_96}. Universal pressure profile (UPP)  for electrons $\mathbb{P}_{e}(r)$
depends on four parameters: the normalization factor $P_0$, the scale radius
$r_p = R_{500}/c_{500}$ (where $c_{500}$ is the concentration parameter for region bounded
by $R_{500}$), and three power indices $a_p$,$b_p$, $c_p$, which denote distribution slope
for different spatial regions ($r<r_s$, $r\approx r_s$ and $r > r_s$, respectively):

 \begin{equation}
    \mathbb{P}_{e}(r) = P_0\cdot
    \left(r/r_p \right )^{-c_p }\cdot(1 + (r/r_p)^{a_p})^{(c_p-b_p)/a_p)}
    \label{eqn:gNFW}
\end{equation}

In order to describe physical processes in particular GC, it is necessary to multiply
UPP (\ref{eqn:gNFW}) by several scale factors:
$P_e(r) = \mathbb{P}_e(r) \times P_{500} \times F_{500},$ where $P_{500}$  is the characteristic pressure \cite{gNFW_07}, \cite{Arnaud_etal_10},
 which can be determined by mass and cosmological parameters:
  \begin{equation}
    P_{500} = 1.65 \times 10^{-3} E(z)^{8/3} \left ( \frac{M_{500}}
    {3 \times 10^{14} h_{70}^{-1} M_{\odot}} \right )^{2/3} h_{70}^2
    \label{eqn:CharP}
\end{equation}

\noindent and the corrective factor $F_{500}$  makes modification by mass due to the self-similarity of GC:

 \begin{equation}
    F_{500} = \left ( \frac{M_{500}}{3 \times 10^{14} h_{70}^{-1} M_{\odot}} \right )^{0.12}
    \label{eqn:CharF}
\end{equation}
\noindent Here $H_0=100h$ km/s/Mpc, $h_{70} = h/0.70$, and $E(z) = H(z)/H(0)$ or
$E^2(z) = \Omega_M(1+z)^3 + \Omega_{\Lambda}$.

For ICM with hydrogen X=0.725, helium Y=0.270 and heavy elements' mass fraction Z = 1-X-Y= 0.005, the mean molecular weights are
$\mu_{gas}=0.60$, $\mu_e=1.15$, and now we can
express the pressure, density and temperature  of ICM gas through the pressure and number
density of ICM electrons \cite{Adam_etal_20}:
 \begin{equation}
    P_{gas}(r) = (\mu_e/\mu_{gas}) P_e(r),\,
     \rho_{gas}(r) = (\mu_e/\mu_{gas})m_H n_e(r),\,
     k_BT_{gas}(r) = \mu_{gas}m_H (P_{gas}(r)/\rho_{gas}(r))
    \label{eqn:GNFW_TotalPressure}
\end{equation}

%\begin{equation}
%    \rho_{gas}(r) = (\mu_e/\mu_{gas})m_H n_e(r)
%   \label{eqn:GasDensity}
%\end{equation}

%\begin{equation}
%    k_BT_{gas}(r) = \mu_{gas}m_H (P_{gas}(r)/\rho_{gas}(r))
%    \label{eqn:GasTemperature}
%^\end{equation}

Magnetic field distribution is following matter distribution \cite{Bonafede_etal_10}, \cite{Murgia_etal_04}.
Based on magnetic freezing conditions, spatial distribution of magnetic field can be expressed as
proportional to some power $\eta_B$ of density profile: $B(r)=B_0(n_e(r)/n_{e,0})^{\eta_B}$.
Normalisation factor  $B_0$ is set to $5 \mu G$, as typical value for GCs, that was measured
for Coma cluster based on Faraday rotation data. \cite{Bonafede_etal_10}.

 Spatially-energetic distribution of CR - nuclei (mainly protons, i=p) and electrons
 ($i=e_1$ for primary ICM electrons, $i=e_2$ for secondary electrons produced in inelastic
 collisions) in ICM can be divided into the product of energetic and spatial distributions:

 \begin{equation}
 dN_i(E,r)/dE=A_i f_i(E) \phi_i(r)
 \label{eqn:CRDistribution}
\end{equation}

ICM protons and electrons, accelerating on shock waves inside GCs, acquire a power
law distribution by energies. In particular, the classic power law with exponential
cut-off energy spectrum  was used in this work:

\begin{equation}
    f_i(E) =  A_i \times \left ( \frac{E}{E_{0,i}} \right )^{-\alpha_i} \times exp \left ( -\frac{E}{E_{cut,i}} \right )
    \label{eqn:CRSpectrum}
\end{equation}

\noindent where $\alpha_i$ is the spectral
index, and cut-off occurs at energies $E_{cut,i}$ ($i=p,e1$).
Normalization factor $A_i$ is determined by the ratio of CR energy to thermal energy inside sphere with radius $r=R_{500}$. This ratio can be calculated as $X_{cr,i,th}= U_{CR,i}(R_{500}) / U_{th}(R_{500})$, where $U_{CR}$ and $U_{th}$ are CR and thermal energy, respectively. Typical values are $X_{cr,p,th} \approx 0.02-0.10$,  $X_{cr,e,th} \approx 0.0001-0.001$.
Spatial distribution of CR in ICM is related to gas distribution and is approximated by power law dependence:

\begin{equation}
    \phi_i(r) =  \left ( \frac{n_e(r)}{n_e(0)} \right)^{\eta_{CR,i}}
    \label{eqn:CRRadialDistr}
\end{equation}

It is also important to note such key parameter as the truncation radius $R_{trunc}$, which is used to denote physical boundaries of region that contains entire GC volume, outside which density drops to zero. There is  discontinuity in thermodynamic parameters at this distance from center of GC. It indicates region of  accretion shock radius, where kinetic energy of accretion is converted into thermal energy \cite{Hurier_etal_19}. Size of this region depends on characteristic radius  $R_{trunc} = 3 R_{500}$.

\section*{\sc gamma-ray and neutrino emission  of A2151}

Characteristics of thermal and non-thermal emission of GCs are determined by
spatial distribution of matter (dark and barionic),
magnetic fields and CR (leptonic and hadronic components) inside ICM.
To recover this distributions we can use observational data about thermal X-ray emission and CMB distortions due to the thermal Sunyaev-Zel’dovich (tSZ) effect.
These data
allow to recover a spatial distribution of partial pressure of electrons $P_e(r)$
and of total pressure of baryonic gas $P_{tot}(r)$ inside ICM. According to presented
above formulae we can build then up the distributions of magnetic fields and CR inside ICM.

Recent work \cite{Tiwari_PalSingh_20} presents new observational results concerning thermal X-ray
emission and recovering of thermodynamic parameters of ICM plasma of brightest subclumps
A2151CB and A2151CF in GC A2151.
 Corresponding parameters of ICM in subclumps A2151CB and A2151CF we used for modeling of their
 gamma-ray and neutrino emission, are presented in Table \ref{tab:InitialParam}.
 Based on these data, we calculated normalization factor $P_0$ for pressure profile from
 Eq. (\ref{eqn:gNFW}). Its value is denoted by star in Table \ref{tab:InitialParam}.
 Other parameters of pressure distribution was taken from Plank Universal
 Profile 2013 \cite{Planck_UPP_13}. We need also
 to take into account the gamma-ray flux absorption by interaction with an
 extragalactic background light (EBL) \cite{Dwek_Krennrich_13}. For this
 purpose we have used Python package
 ebltable\footnote{https://github.com/me-manu/ebltable/}.

Using described above parameters of galaxy cluster and cosmic rays  as input ones,
we use open available online\footnote{https://github.com/remi-adam/minot}
Python language-based   software  MINOT
(modeling the intracluster medium (non-)thermal content and observable prediction
tools). MINOT simulates the intracluster diffuse thermal and nonthermal broad-band
(from radio- to gamma-ray)  and neutrino  emission and as an output calculates
the multiwavelength observable spectra, profiles, fluxes, and images for  the radio
(synchrotron),  X-ray (thermal Bremsstrahlung), gamma-ray (inverse Compton and
hadronic processes), and  neutrino (hadronic processes) messengers \cite{Adam_etal_20}.

\begin{table}[h]
 \centering
 \caption{Simulation parameters }\label{tab:InitialParam}
 \vspace*{1ex}
 \begin{tabular}{ccc}
  \hline
  \bf{Global Parameters} & A2151CB & A2151CF \\
  \hline
  \hline
  z & 0.0368 & 0.0368 \\
 \hline
  $M_{500}$ in $[ 10^{13} M_{\odot}]$ & $ 9.08 \pm 5.24 $ & $3.01 \pm 2.11$ \\
 \hline
 $R_{500}$ in $[kpc]$ & $ 803.38\substack{+113.18 \\ -172.13}$ & $ 566.09\substack{+89.09 \\ -158.26} $ \\
 \hline
 $R_{trunc}$ in $[kpc]$ & 2410.15 & 1698.28 \\
 \hline
 helium mass fraction & 0.2735 & 0.2735 \\
 \hline
 metallicity & 0.0153 & 0.0153 \\
 \hline
 abundance & 0.43 & 0.13 \\
 \hline
 hydrostatic mass bias & 0.2 & 0.2 \\
 \hline
  $X_{cr,p}=U_{cr,p}/U_{th}$ inside $R_{500}$ & 0.1 & 0.1 \\
 \hline
  $X_{cr,e1}=U_{cr,e1}/U_{th}$ inside $R_{500}$ & 0.01 & 0.01 \\
 \hline
  $E_{p,min}$ in [GeV] & 1.21 & 1.21 \\
 \hline
  $E_{p,cut}$ in [TeV] & $30$ & $30$ \\
   \hline
  $E_{p,max}$ in [TeV] & $10^6$ & $10^6$ \\
 \hline
 Spectral index $\alpha_{p}$  (Model)  &  1.5 (PLEC)  & 1.5 (PLEC) \\
 \hline
 $E_{e1,min}$ in [keV] & 511 & 511 \\
 \hline
 $E_{e1,break}$ in [GeV] & - & - \\
 \hline
  $E_{e1,cut}$ in [TeV] & $10^6$ & $10^6$ \\
 \hline
  $E_{e1,max}$ in [TeV]  & $10^6$ & $10^6$ \\
 \hline
 Spectral index $\alpha_{e1}$ (Model) &  2.3  (PL)  & 2.3 (PL) \\
% \hline
% Spectrum CRe & PL, Pivot = 1 TeV, $\alpha = 3.0$ & same value & same %value  \\
 \hline
  Density $CR_p$ model & & \\
    $n_{CR,p}(r)\propto n_e(r)^{\eta_{CR,p}}$ & $\eta_{CR,p}=1.0$  & $\eta_{CR,p}=1.0$ \\
  \hline
  Density $CR_{e1}$ model & & \\
  $n_{CR,e}(r)\propto n_e(r)^{\eta_{CR,e}}$ & $\eta_{CR,e}=1.0$  & $\eta_{CR,e}=1.0$ \\
  \hline
  \hline
 \bf{Pressure gas model (GNFW):} & $P_0$+PUP  & $P_0$+PUP \\
 \hline
 \hline
 $P_0$ $[keV/cm^3]$ & $0.009^{*}$ & $0.0077^{*}$ \\
 \hline
 $a_p$ & 1.33 & 1.33 \\
 \hline
 $b_p$ & 4.13 & 4.13 \\
 \hline
 $c_p$ & 0.31 & 0.31 \\
 \hline
 $c_{500}$ & 1.81 & 1.81 \\
 \hline
 \hline
 \bf{Density gas model (beta )} & & \\
 \hline
 \hline
 $n_0$ $[cm^{-3}]$ & 0.00875 & 0.0049\\
 \hline
 $r_c$ [kpc] & 19.35 & 10.43 \\
 \hline
 $\beta_{dens}$ & 0.38 & 0.28 \\
 \hline
 \hline
 \bf{Magnetic field (Eq.(\ref{eqn:beta-magfield}))} & & \\
 \hline
 \hline
 $B_0$ $[\mu G]$        & 5         & 5 \\
 \hline
 $\eta_B$           & 2/3     & 2/3 \\
 \hline
 \hline

 \end{tabular}
\end{table}

\begin{figure}[!h]
\centering
\begin{minipage}[t]{.48\linewidth}
\centering
\epsfig{file = 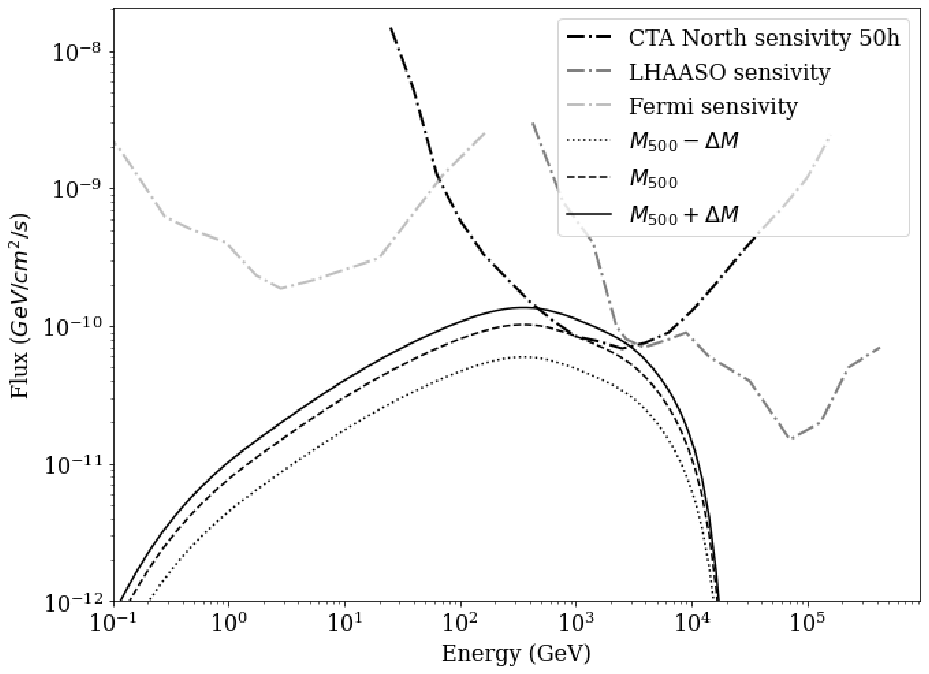,width = .85\linewidth}
\caption{Spectral flux from A2151CB for three values of cluster mass $M_{500}$ and $M_{500} \pm \Delta m$. Other parameters as in Table \ref{tab:InitialParam}}\label{fig:A2151CBFluxDifMass}
\end{minipage}
\hfill
\begin{minipage}[t]{.48\linewidth}
\centering
\epsfig{file = 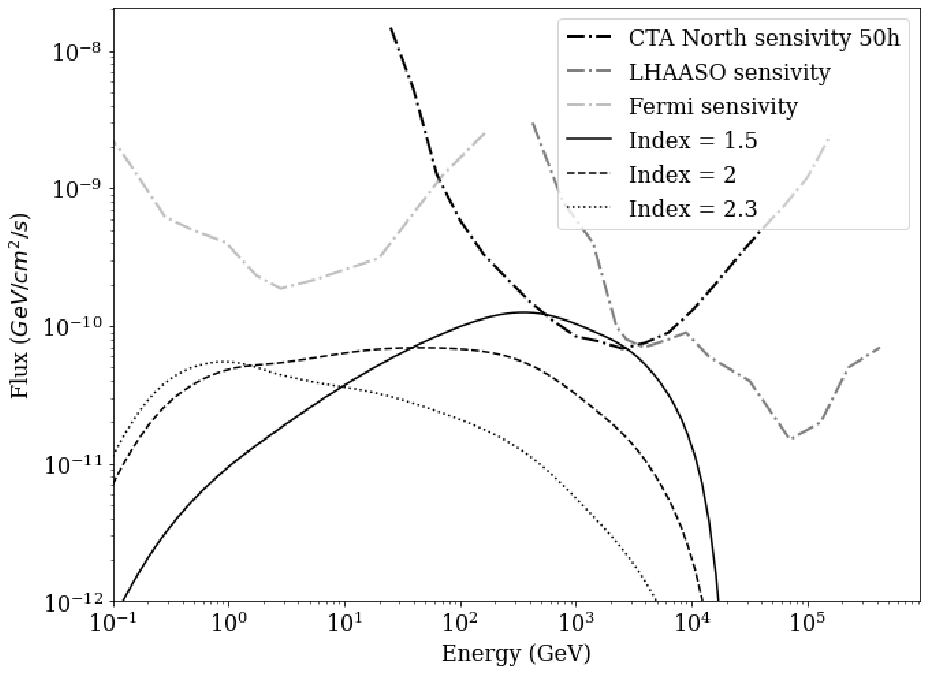,width = .85\linewidth}
\caption{Spectral flux from A2151CB for three values of spectral index. CRs distribution $CRp \propto n_e(r)^2$. Other parameters as in Table \ref{tab:InitialParam}}\label{fig:A2151CBFluxDifIndexCRP2}
\end{minipage}
\hfill
\begin{minipage}[t]{.48\linewidth}
\centering
\epsfig{file = 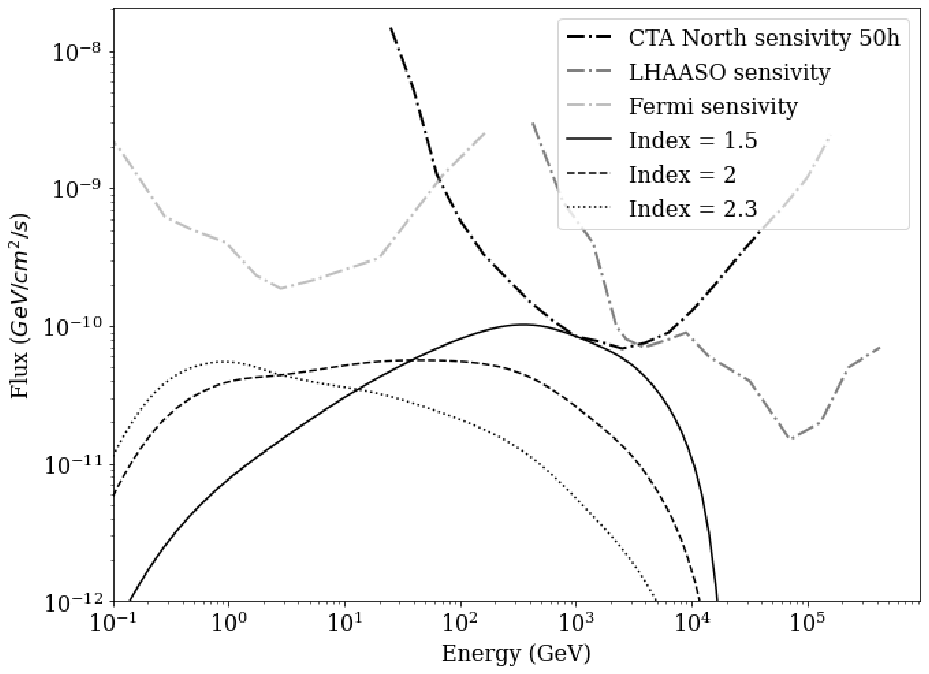,width = .85\linewidth}
\caption{Spectral flux from A2151CB for three values of spectral index. Other parameters as in Table \ref{tab:InitialParam}}\label{fig:A2151CBFluxDifIndex}
\end{minipage}
\hfill
\begin{minipage}[t]{.48\linewidth}
\centering
\epsfig{file = 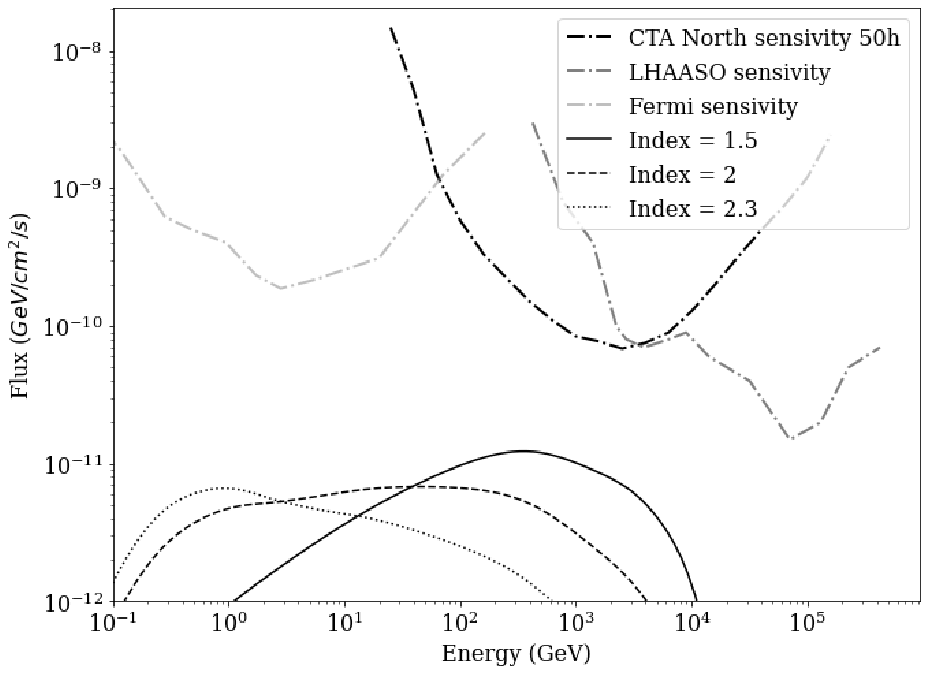,width = .85\linewidth}
\caption{Spectral flux from A2151CF for three values of spectral index. Other parameters as in Table \ref{tab:InitialParam}}\label{fig:A2151CFFluxDifIndex}
\end{minipage}
\end{figure}

Main results of gamma-ray and neutrino emission for A2151CB and A2151CF subclumps are presented in Fig 1-5. The most promising for CTA-detection cases correspond to hard CR spectrum (spectral index $\alpha_p<2$).  In particular, Fig. \ref{fig:A2151CBFluxDifMass} demonstrates expected gamma-ray flux from the central brightest subclump A2151CB for $\alpha_p=1.5$  and for three  values of cluster mass - average mass and $\pm \Delta m$ from Table \ref{tab:InitialParam}.
Meantime, there is no significant increase in flow due to more compact CR distribution (a result of magnification of $\eta_{CR,p}$ twice is shown in Fig. \ref{fig:A2151CBFluxDifIndexCRP2}).

\begin{figure}[!h]
\centering
\begin{minipage}[t]{.48\linewidth}
\centering
\epsfig{file = 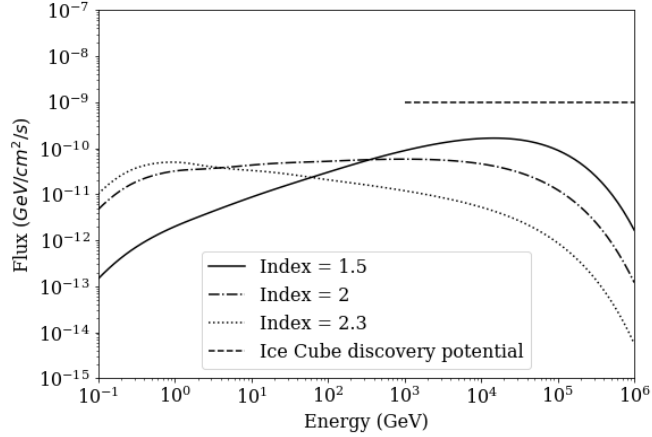,width = 1\linewidth}
\caption{Neutrino flux from the brighest subclump A2151CB with IceCube $5 \sigma$ discovery potential at 1-1000 TeV \cite{IceCube_potential_19}}

\label{fig:NeutrinoFlux}
\end{minipage}
\end{figure}

For comparing different models of CR distribution in ICM, we have simulated gamma-ray flux for three values of spectral index - see  Fig. \ref{fig:A2151CBFluxDifIndex} and Fig. \ref{fig:A2151CFFluxDifIndex}.  As we can see from these figures, there is a possibility for CTA to detect the brighest subclump A2151CB in case of hard CR spectrum, instead of A2151CF, that has relatively low flux for all CR distribution models (Fig. \ref{fig:A2151CFFluxDifIndex}.
Calculated neutrino flux  from the brightest subclump A2151CB together with IceCube discovery potential at significance level 5$\sigma$ is shown in Fig. \ref{fig:NeutrinoFlux}. As we can see from this figure, neutrinos do not suffer from energy losses as photons do due to EBL absorption. This fact can make neutrino emission to be a good probe for GC in the future.

\section*{\sc  conclusions}

 Due to a long  time for cosmic rays diffusion escape, GCs should  be a luminous extragalactic sources
 of  neutrino and non-thermal gamma-rays, but  such emission has not been detected yet.
 Using the recent data  from \cite{Tiwari_PalSingh_20}
 about thermal X-ray emission of brightest subclumps A2151CB and  A2151CF in GC A2151 we  have carried
 out a modelling of non-thermal hadronic gamma-ray and neutrino emission of these subclumps using
 MINOT code \cite{Adam_etal_20}. Results of our simulations show that  under the typical parameters
 of cosmic ray distribution (cosmic ray proton to thermal energy ratio $X_{cr.p}\approx 0.04-0.06$,
 spectral index of proton power-law energy spectrum $\gamma=2.5$)  both  non-thermal hadronic
 gamma-ray and neutrino  emission of  subclumps A2151CB and  A2151CF are still undetected by
 existing and planned detectors. Meantime, backreaction of 7 AGN-type galaxies in subclumps
 A2151CB and A2151CF and other unaccounted for sources of cosmic rays (merging and accretion
 flows, turbulence etc.) can  increase $X_{cr.p}\approx 0.1$ and  provide harder spectral index
$\gamma=1.5-2.0$. In these cases brighter subclump A2151CB can be detected by CTA at
$5 \sigma$ level. Hard cosmic ray proton spectrum is promising also for detection of neutrino fluxes
from GCs in future neutrino detectors (IceCube-Gen2 etc.). Contrary to gamma-ray fluxes, neutrino fluxes do not attenuate by interaction with the  extragalactic background light (EBL), therefore,
there are promising possibilities to detect PeV- neutrinos in case of GCs - cosmic PeVatrons with hard ($\gamma < 2.0$) spectra.

\end{document}